\documentclass[5p,authoryear,preprint]{elsarticle}

\usepackage[breaklinks,colorlinks]{hyperref}
\newcommand{\epkobs}{\ensuremath{E_{pk}^{(obs)}}}
\newcommand{\xpkobs}{\ensuremath{X_{pk}^{(obs)}}}

\newcommand{\xs}{\ensuremath{X_S}}

\newcommand{\tobs}{\ensuremath{t_{jet}^{(obs)}}}
\newcommand{\xtobs}{\ensuremath{X_{t_{jet}}^{(obs)}}}

\newcommand{\dobs}{\ensuremath{T_{45}^{(obs)}}}
\newcommand{\xdobs}{\ensuremath{X_{45}^{(obs)}}}

\newcommand{\epksrc}{\ensuremath{E_{pk}^{(src)}}}
\newcommand{\xpksrc}{\ensuremath{X_{pk}^{(src)}}}
\newcommand{\eiso}{\ensuremath{E_{iso}}}
\newcommand{\xiso}{\ensuremath{X_{iso}}}
\newcommand{\tsrc}{\ensuremath{t_{jet}^{(src)}}}
\newcommand{\xtsrc}{\ensuremath{X_{t_{jet}}^{(src)}}}
\newcommand{\dsrc}{\ensuremath{T_{45}^{(src)}}}
\newcommand{\xdsrc}{\ensuremath{X_{45}^{(src)}}} %
\newcommand{\vxsrc}{\ensuremath{{\mathbf x}^{(src)}}}
\newcommand{\vxobs}{\ensuremath{{\mathbf x}^{(obs)}}}
\newcommand{\vA}{\ensuremath{\mathbf A}}
\newcommand{\vx}{\ensuremath{\mathbf x}}
\newcommand{\vu}{\ensuremath{\mathbf u}}
\newcommand{\vv}{\ensuremath{\mathbf v}}
\newcommand{\vB}{\ensuremath{\mathbf B}}

\newcommand{\vy}{\ensuremath{\mathbf y}}
\newcommand{\vL}{\ensuremath{\mathbf L}}
\newcommand{\vf}{\ensuremath{\mathbf f}}
\newcommand{\vn}{\ensuremath{\mathbf n}}
\newcommand{\vw}{\ensuremath{\mathbf w}}
\newcommand{\vg}{\ensuremath{\mathbf g}}
\newcommand{\ve}{\ensuremath{\mathbf e}}
\newcommand{\vb}{\ensuremath{\mathbf b}}
\newcommand{\vD}{\ensuremath{\mathbf D}}
\newcommand{\vQ}{\ensuremath{\mathbf Q}}
\newcommand{\vs}{\ensuremath{\mathbf s}}

\journal{New Astronomy}

\begin{document}

\begin{frontmatter}

\title{GRBs as standard candles: There is no ``circularity problem''\\(and
there never was)}
\author[cg]{Carlo Graziani\corref{cor1}}
\ead{carlo@oddjob.uchicago.edu}
\cortext[cor1]{Corresponding Author}
\address{Department of Astronomy \& Astrophysics, University of Chicago,
5640 S. Ellis Avenue, Chicago, IL 60637, USA.
Tel: +01-773-702-7973 ; FAX: +01-773-702-6645}


\begin{abstract}

Beginning with the 2002 discovery of the ``Amati Relation'' of GRB spectra,
there has been much interest in the possibility that this and other
correlations of GRB phenomenology might be used to make GRBs into standard
candles.  One recurring apparent difficulty with this program has been that
some of the primary observational quantities to be fit as ``data'' --- to
wit, the isotropic-equivalent prompt energy $\eiso$ and the
collimation-corrected ``total'' prompt energy energy $E_{\gamma}$ ---
depend for their construction on the very cosmological models that they are
supposed to help constrain.  This is the so-called ``circularity problem''
of standard candle GRBs.  This paper is intended to point out that the
circularity problem is not in fact a problem at all, except to the extent
that it amounts to a self-inflicted wound.  It arises essentially because
of an unfortunate choice of data variables --- ``source-frame'' variables
such as $\eiso$, which are unnecessarily encumbered by cosmological
considerations.  If, instead, the empirical correlations of GRB
phenomenology which are formulated in source-variables are {\it mapped to
the primitive observational variables} (such as fluence) and compared to
the observations in that space, then all taint of circularity disappears. 
I also indicate here a set of procedures for encoding high-dimensional
empirical correlations (such as between $\eiso$, $\epksrc$, $\tsrc$, and
$\dsrc$) in a ``Gaussian Tube'' smeared model that includes both the
correlation and its intrinsic scatter, and how that source-variable model
may easily be mapped to the space of primitive observables, to be convolved
with the measurement errors and fashioned into a likelihood.  I discuss the
projections of such Gaussian tubes into sub-spaces, which may be used to
incorporate data from GRB events that may lack some element of the data
(for example, GRBs without ascertained jet-break times).  In this way, a
large set of inhomogeneously observed GRBs may be assimilated into a single
analysis, so long as each possesses at least two correlated data
attributes.

\end{abstract}

\end{frontmatter}

\section{Introduction\label{Sec:Intro}}

Since the earliest published evidence of tight correlations in gamma-ray
burst (GRB) spectral properties \citep{amati_etal_2002}, there has been
sustained interest in pressing those correlations into service to make GRBs
into standard candles, which is the same office that the Phillips
correlation performs for SN~Ia
\citep{phillips_1993,riess_etal_1998,goldhaber_perlmutter_1998}. The
intriguing possibility is that GRBs may open a window in redshift space
($z\sim[1-8]$) beyond what is provided by SN~Ia studies, for the purpose of
constraining the parameters that characterize Dark Energy
\citep{dai_etal_2004,ggl_2004,friedman_bloom_2005,liang_zhang_2005,firmani_etal_2005}.

The earliest correlation, the ``Amati Relation'', discovered by
\citet{amati_etal_2002}, was between the isotropic-equivalent prompt
emission energy $\eiso$ and the peak energy $\epksrc$ of the Band-function
spectrum fit to the time-integrated prompt emission from the burst, boosted
to the source frame by the expansion factor $1+z$.  Other correlations
were discovered in short order, ostensibly exhibiting tighter scatter that
could make them more suitable for standardizing candles.  Examples are the
``Ghirlanda Relation'' \citep{ggl_2004}, connecting the
collimation-corrected prompt energy $E_\gamma$ and $\epksrc$; the
``Liang-Zhang Relation'' \citep{liang_zhang_2005}, connecting $\eiso$ with
a fit-determined function constructed from $\epksrc$ and the source-frame
jet-break time $\tsrc$; and the ``Firmani Relation''
\citep{firmani_etal_2006}, analogous to the Liang-Zhang relation, but
replacing the dependence on $\tsrc$ with one on $\dsrc$, the source-frame
``emission time,'' which is a duration measure that robustly stands up to the
diversity of duty cycles observed in prompt GRB emission
\citep{reichart_etal_2001}.

The later correlations of \citet{liang_zhang_2005} and
\citet{firmani_etal_2006} constitute considerable advances over the earlier
work on constructing GRB distance indicators.  By passing from
($E_\gamma,\epksrc$) to a 2-D projection of the space
$(\eiso,\epksrc,\tsrc)$, \citet{liang_zhang_2005} eliminated all reference
to highly uncertain theoretical factors --- the density of the ISM in the
burst source neighborhood, and the conversion efficiency of kinetic energy
to radiation in the afterglow --- required to convert $\tsrc$ to a jet opening
angle. This purged an important source of systematic error from the
problem.  \citet{firmani_etal_2006} went a step further, passing to
2-D projections of the space $(\eiso,\epksrc,\dsrc)$, which, by
replacing the difficult-to-obtain jet-break time with the more easily
measured prompt duration made many more GRBs available as potential
standard candles.

A fly in the ointment was noticed early on by several authors
\citep{friedman_bloom_2005,liang_zhang_2005,firmani_etal_2005}: The Amati
and Ghirlanda relations were calibrated assuming a standard concordance
$\mathnormal{\Lambda}$CDM cosmology.  That is to say, it is not possible to construct
quantities such as $\eiso$ or $E_\gamma$ from the observed GRB prompt
fluence $S$ without reference to a specific cosmological model to supply
the luminosity distance.  As the cosmological model is precisely what is to
be constrained from the data, an inconsistency would appear to have been
introduced into the problem.  This is the (apparent) ``Circularity
Problem'' of GRB standard candles.

Much effort and ingenuity has gone into the abatement of the circularity
problem.  \citet{friedman_bloom_2005} performed fits of the Ghirlanda
Relation assuming a wide range ($0\le\mathnormal{\Omega}_M,\mathnormal{\Omega}_\mathnormal{\Lambda}\le2$) of
``fiducial'' cosmologies, and used each such fit to infer confidence
regions on the true parameters, reporting regions that bracket all the
results.  Aside from being rather conservative, it is difficult to
understand what sort of confidence probability is to be ascribed to such
intervals.  This is problematic if confidence intervals from GRB studies
are to be combined with those from other types of Dark Energy studies, such
as SN~Ia.

The procedures adopted by \citet{liang_zhang_2005} and by
\citet{firmani_etal_2005} are, from a conceptual point of view, even more
problematic.  \citet{liang_zhang_2005} re-fit the correlation for each
``fiducial'' cosmology, to obtain a $\chi_{corr}^2$.  For each such fit to
the correlation, they fit to the cosmological parameters, and re-weight the
probability of the cosmological parameter fit by $\exp(-\chi_{corr}^2/2)$
--- in effect, an ad hoc, tacit, and oddly data-dependent choice of prior.

\citet{firmani_etal_2005} explicitly embrace Bayesian logic, by
interpreting the likelihood obtained for cosmology $\mathnormal{\Omega}$ using
Ghirlanda-relation fits obtained assuming ``fiducial'' cosmology
$\bar\mathnormal{\Omega}$ as a conditional probability $P(\mathnormal{\Omega}|\bar\mathnormal{\Omega})$ --- an
interpretation that is both mathematically inconsistent (such an expression
should be proportional to the Dirac delta function
$\delta(\mathnormal{\Omega}-\bar\mathnormal{\Omega})$), and logically dubious (what information could
cosmology $\bar\mathnormal{\Omega}$ possibly supply about cosmology $\mathnormal{\Omega}$?)  They
then eliminate $\bar\mathnormal{\Omega}$ from their results by marginalizing this
probability with some prior on $\bar\mathnormal{\Omega}$.  This removes the nuisance
parameter $\bar\mathnormal{\Omega}$ from the final expressions, but does not correct the
logical inconsistency that underlies the calculation.

More recently, an ``astronomical'' fix has been proposed for the
circularity problem.  \citet{liang_etal_2008} interpolate distance moduli
from SN~Ia at the same redshift ($z<1.4$) to ``train'' the correlations at
low redshift.  This is not terribly different from using nearby SN~Ia to
calibrate the Phillips relation for all SN~Ia, and is not conceptually as
problematic as some of the above approaches.  However it is a rather weak
solution, since the relation must be calibrated using a small subset of
GRBs, and since it means that GRB distance moduli can never even in
principle be determined more accurately than SN~Ia distance moduli. 
Moreover, if confidence regions on Dark Energy parameters obtained using
such a calibration are to be combined with confidence regions obtained from
SN~Ia, a new hidden statistical dependence will have been introduced that
will be difficult to characterize.

It is unfortunate that so much effort has been thus addressed to solving
this problem.  As I show below, {\it there is no real circularity problem},
and there never was.  To the extent that a ``problem'' exists, it is, in
effect, a self-inflicted wound, arising from an unfortunate choice of data
variables --- ``source-frame'' variables such as $\eiso$ and $E_\gamma$,
which are, by their construction, unnecessarily encumbered by cosmological
considerations.  If, instead, the empirical correlations of GRB
phenomenology which are formulated in source-variables are {\it mapped to
the primitive observational variables} such as fluence (so that the model
may discharge its duty of predicting the data, without at the same time
being obliged to assist in constructing it), then the circularity
disappears, and the analysis may be carried out without fear of
inconsistency or paradox.

A recent paper by \citet{basilakos_etal_2008} addresses the circularity
issue by making explicit the dependence of ``data'' such as $\eiso$ or
$E_\gamma$ on the cosmological parameter $\mathnormal{\Omega}_M$ in the expression for
the log-likelihood, and allows both the ``data'' and the model to vary with
the model parameters in the fit.  As such, this work does not make a clean
separation between model and data, in the way that is in my opinion
desirable.  Nonetheless, for reasons that will be discussed in
\S\ref{subsec:GaussianDensity}, the resulting formalism has some features
that are similar to the one presented here.

Concomitantly with the necessary disentangling of data from cosmological
modeling, I show below how multi-dimensional correlations of the sort
projected down to two dimensions by \citet{liang_zhang_2005} and
\citet{firmani_etal_2006} can be fully, and more informatively, modeled in
the higher-dimensional space in which they reside, by a {\it Gaussian Tube}
model, which represents the correlation together with its intrinsic
scatter.  The Gaussian nature assumed for the scatter yields the benefit of
easy convolution with measurement errors to furnish a likelihood function
that may be put to the usual inferential work.  The Gaussian Tube will be
illustrated in this work by formulating it in the 4-D space of
source-frame variables ($\tsrc$, $\dsrc$, $\epksrc$, $\eiso$), and mapping it to the
space of observables ($\tobs$, $\dobs$, $\epkobs$, $S$).  Generalization to
higher-dimensional spaces or to other observables is obvious and immediate.

For those GRBs that are endowed with all four observations, the full
Gaussian Tube model is used to produce the event likelihood ${\cal L}_i$. 
For GRBs that are missing some measured observables, we may still calculate
an event likelihood by using the projection of the Tube model into the
space of available observables, whether that be 2-D or 3-D (the projection
onto 1-D is a uniform distribution, which is uninformative).  Thus it is
possible to fit simultaneously to {\it all} GRBs for which at least
two correlated observables are measured.  This is a substantial technical
advance, in that it was previously necessary to use separately samples of
GRBs with different available measurements.

Projection of the tube model has additional uses beyond extending the data
set.  A mysterious multi-dimensional tube correlation model, however
technically satisfying, is not persuasive unless one can verify that the
data in fact justify the model.  Fortunately, this is not hard to do.  Once
a best-fit cosmology $\mathnormal{\Omega}$ has been obtained (or once we have fixed
$\mathnormal{\Omega}$ at the concordance model), we may project the best-fit Gaussian
Tube model into various 2-D planes --- exhibiting both its orientation and
its Gaussian width --- and superpose the applicable data in that plane,
including measurement errors.  We are thus able to exhibit the various
existing 2-D correlations as different perspectives on the full,
multi-dimensional correlation in a series of 2-D plots, and visually
inspect the agreement with the data.

The organization of the remainder of this paper is as follows:  in
\S\ref{Sec:TubeModel} I introduce the variables in play, define some
notation, and exhibit the Gaussian Tube model in technical detail.  In
\S\ref{Sec:Proj}, I discuss the mathematical details associated with
projection operations of the model into lower-dimensional spaces.  In
\S\ref{Sec:ModelData} I discuss the procedures required to compare the
model to data --- formulation of the event likelihoods for the cases of
full and partial data, and how the event likelihoods are (trivially) strung
together into a full likelihood function for the ensemble of GRBs.  In
\S\ref{Sec:Sanity} I discuss the use of model projections to verify that
the data in fact has a nodding acquaintance with the
difficult-to-visualize, multi-dimensional model.  A discussion of likely
data requirements of the method presented here is in
\S\ref{Sec:DataRequirements}. Final discussion and conclusions appear in
\S\ref{Sec:Discussion}.

\section{The Gaussian Tube Correlation Model\label{Sec:TubeModel}}

The Gaussian Tube is defined as a density which is Gaussian about a
symmetry axis along the direction of the correlation, and invariant along
that axis.  The finite-width density is intended to represent the intrinsic
scatter of the correlation. The model is a somewhat crude empirical
approximation, since it does not allow for the nature of the intrinsic
scatter in the correlation to change as one moves up or down the symmetry
axis.  The benefit of the simplification is that the likelihood function of
data endowed with Gaussian measurement errors may be computed analytically,
as I will show in \S\ref{Sec:ModelData}.  Some possibilities for moving
beyond this simplification are indicated at the end of
\S\ref{Sec:Sanity}.

It is convenient to work with the logs of the observables as
primary quantities.  Accordingly, we define
\begin{eqnarray}
\xtsrc\equiv\log\tsrc &;& \xdsrc\equiv\log\dsrc ;\nonumber\\
\xiso\equiv\log\eiso &;& \xpksrc\equiv\log\epksrc ;\nonumber\\
\xtobs\equiv\log\tobs &;& \xdobs\equiv\log\dobs ;\nonumber\\
\xs\equiv\log S &;& \xpkobs\equiv\log\epkobs,
\end{eqnarray}
and introduce the vectors
\begin{equation}
{\vxsrc}\equiv\left[
\begin{array}{c}
\xtsrc\\\xdsrc\\\xiso\\\xpksrc
\end{array}
\right]
\end{equation}
\begin{equation}
{\vxobs}\equiv\left[
\begin{array}{c}
\xtobs\\\xdobs\\\xs\\\xpkobs
\end{array}
\right]
\end{equation}
\begin{equation}
{\vf(z,\mathnormal{\Omega})}\equiv
\left[
\begin{array}{c}
\log(1+z)\\\log(1+z)\\-\log[4\pi(1+z)^{-1} d_L(z,\mathnormal{\Omega})^2]\\-\log(1+z)
\end{array}
\right],
\end{equation}
where $z$ is the redshift of a particular GRB.  In terms of this notation,
the transformation from source variables to primitive observables of a
particular GRB is simply
\begin{equation}
\vxobs = \vxsrc + \vf(z,\mathnormal{\Omega}).
\label{src2obs}
\end{equation}
The transformation is thus an elementary shift, albeit one that is
different for each GRB (since each is at a different redshift $z$).

We will define the model in the space of $\vxsrc$, and use this relation to
move it to the observable space when the time comes to compare the model to
data.

\subsection{The Axis Of The Tube}

The symmetry axis of the event density distribution is easily defined in
terms of elementary analytical geometry. The direction of of the tube axis
is along a vector $\vn$, and the axis passes through a point $\vx_0$, so
that points on the axis are defined by the parametric relation
$\vx=\vx_0+t\vn$, for all real $t$.

This parametrization is not unique, since $\vn$ may be multiplicatively
rescaled, and $\vx_0$ may be shifted by a multiple of $\vn$.  In order to
fix a non-degenerate parametrization it is necessary to choose a definite
scale for $\vn$ and a definite intercept for $\vx_0$.  We will choose
$\vn^T=[n_1,\,n_2,\,n_3,\,1]$\footnote{
The more familiar scale choice of $\vn\cdot\vn=1$ (i.e. choosing a unit
vector for $\vn$) is not particularly natural in this context.  The reason
is that there is no natural Euclidean metric defined in the space of
observables, and we have no particular reason to import one.  The cost of
the additional complexity introduced by a quadratic normalization
convention is not offset by any benefit (such as, for example, a
normalization that is invariant under a relevant class of reparametrizations). 
}, and $\vx_0^T=[x_{0,1},\,x_{0,2},\,x_{0,3},\,0]$.  Thus 6 parameters are
required to specify the axis.

\subsection{The Gaussian Density\label{subsec:GaussianDensity}}

The Gaussian Tube density is denoted by $\rho(\vxsrc)\,d^4\vxsrc$, where
\begin{eqnarray}
\rho(\vxsrc)&=&N\times\nonumber\\
&&\hspace{-1.5cm}\exp\left[
-\frac{1}{2} \left(\vxsrc-\vx_0\right)^T\vB
\left(\vxsrc-\vx_0\right)
\right].\nonumber\\
\label{gt_density}
\end{eqnarray}
and where $\vB$ is a non-negative-definite matrix.

The eigenvectors of $\vB$ are the principal directions of the ellipsoids of
constant density.  If the eigenvalue corresponding to a certain eigenvector
should become very small, the ellipsoids will become very elongated along
the corresponding direction.  In the limit of an eigenvalue going to zero,
the distribution will be infinitely elongated, becoming, in effect, a
tube.  The condition that the tube should be oriented along the direction
$\vn$ is therefore $\vB\cdot\vn=0$.

We require a useful parametrization of $\vB$ that will satisfy this
condition.  Such a parametrization may be exhibited by introducing dual
vectors (linear maps from vectors to numbers) $\vw_i$, $i=1,2,3$,
satisfying $\vw_i(\vn)=0$.  Then, we may choose
\begin{equation}
\vB=\sum_{i,j=1}^3b^{ij}\vw_i\vw_j,
\label{B1}
\end{equation}
where the $b^{ij}$ are components of a positive-definite matrix.  By
construction, this $\vB$ evidently satisfies $\vB\cdot\vn=0$.

A convenient choice of the $\vw_i$ may be specified in terms of the dual
basis $\vg_\nu$, $\nu=1,\ldots,4$, which is dual to the coordinate
direction vectors $\ve^\mu$, $\mu=1,\ldots,4$, in the sense that
$\vg_\nu(\ve^\mu)=\delta_\nu^\mu$.  Then we may choose
\begin{equation}
\vw_i=\vg_i-n_i\vg_4.
\label{dualbasis}
\end{equation}
It is straightforward to verify that $\vw_i(\vn)=0$ (recall that $n_4=1$ by
convention).  The $\vw_i$ may be written in component form
$\vw_i=\sum_{k=1}^4{w_i}^k\vg_k$, where from Eq.~(\ref{dualbasis})
\begin{equation}
{w_i}^k=\delta_i^k - n_i\delta_4^k.
\label{dualbasis_comp}
\end{equation}

By substituting the components of the $\vw_i$ from
Eq.~(\ref{dualbasis_comp}) into Eq.~(\ref{B1}), we may obtain the matrix
components of $\vB$ along the coordinate dual basis $\vg_\nu$ (which is
what we mean by the ``matrix'' $\vB$):
\begin{equation}
[\vB]^{km}=\sum_{i,j=1}^3b^{ij}{w_i}^k{w_j}^m.
\label{reduced_b}
\end{equation}

It remains to guarantee that the components $b^{ij}$ produce a matrix $\vB$
satisfying $\vx^T\cdot\vB\cdot\vx\ge 0$ for all vectors $\vx$, and
$\vx^T\cdot\vB\cdot\vx=0$ only when $\vx\propto\vn$.  From Eq.~(\ref{B1}),
this is clearly equivalent to the condition that the matrix $\vb$ with
components $b^{ij}$ should be a positive-definite matrix.  A
parametrization that guarantees this is the {\it Cholesky Decomposition}
$\vL$ of $\vb$ \citep[see, e.g.][p. 141]{golub_vanloan_1989}.  This is the
unique lower-triangular matrix with components satisfying
$L_{ii}>0$ and $L^{ij}=0,~j>i$, in terms of which $\vb=\vL\,\vL^T$, that is,
\begin{equation}
b^{ij}=\sum_{n=1}^3 L^{in}L^{jn}.
\label{choldec}
\end{equation}

We therefore adopt the $(3\times 4)/2=6$ components $L^{ij}$ of $\vL$ as 
the parameters which, together with $\vn$, control the quadratic form
$\vB$.

It is necessary at this point to be more definite about the normalization
``constant'' $N$ that figures in Eq.~(\ref{gt_density}).  This
normalization is of course constant with respect to the variables $\vx$. 
It is {\it not} constant with respect to the parameters $\vL$, however.
This is because we must require that the model be somehow normalized, so
that we can vary the shape of the tube (the $\vL$) without varying the
predicted overall rate of GRB events.  This is an essential feature of
the model, without which the task of inferring the $\vL$ from the data
will certainly fail.\footnote{
The failure would take the form of an inability of the likelihood function
to prefer tight correlations to dispersed ones.  Since the term in the
exponential of the Gaussian is negative quadratic, and hence bounded above
by zero, the fit to the data could simply proceed by making $\vB\rightarrow
0$ (which makes the model a uniform density), reaching the maximum
attainable likelihood irrespective of how bad the correlation really
is.  It is the job of the normalization ``constant'' to prevent this
catastrophe.  The normalization of Eq.~(\ref{norm}) guarantees that the
likelihood will decline to zero if $\vB$ attempts to go to zero.
}

The normalization must have the following property: the integral of
$\rho(\vx)$ on any 3-D hyperplane must be independent of $\vL$.  Loosely
speaking, this guarantees that changing the width and ``cross-sectional
shape'' of the Gaussian Tube does not change the overall event rate of
predicted GRBs.  This allows us to decouple the aspects of the model that
predict the correlation shape (which we care about) from the aspects that
predict GRB event rates (which we do not).

This normalization is easily exhibited: it is
\begin{equation}
N=\prod_{i=1}^3L^{ii}, 
\label{norm}
\end{equation}
which is just the square root of the determinant of the matrix $\vb$.  This
is roughly speaking 1/(product of $\sigma$ over non-degenerate principal
directions), which is the standard normalizing factor of Gaussian
distributions (except for inessential $\pi$-related factors, which
fortunately {\it are} constant).

At this point, enough of the model is in view to allow a comparison with
the work of \citet{basilakos_etal_2008}, who consider separately various
2-D correlations.  As pointed out in \S\ref{Sec:Intro}, there was
no clean separation made between model and data in that paper. 
Nonetheless, \citet{basilakos_etal_2008} also work with log-space
observables, so that the relation between their source- and observer-frame
variables is still given by an offset, as in Eq.~(\ref{src2obs}).  Since
they use $\chi^2$, which is a function of model-data difference, it is
immaterial from the point of view of the formula whether the offset is
applied to the model, as here, or the negative offset is applied to the
data, as in their work.  Note, however, that while $\chi^2$ is effectively
the argument of the exponential of the likelihood in
Eq.~(\ref{gt_density}), it does not represent the dependence of the
likelihood on the parameters through the normalization $N$, which, as
argued above, is an important omission.  The effect of the omission may be
seen from Eq.~(5) of \citet{basilakos_etal_2008}, where in the limit of the
slope parameter $a\rightarrow\infty$, the expression for $\chi^2$ saturates
at a constant value, leading to open confidence contours.  It is precisely
this circumstance that the normalization of Eq.~(\ref{norm}) avoids.

\subsection{Sampling From A Gaussian Tube}

Methods for sampling from the distribution defined by a Gaussian Tube are
obviously of some interest if one intends to {\it simulate} events from
such a distribution.  Sampling from the tube is not a straightforward
exercise in multidimensional Gaussian sampling, as one might imagine upon
contemplating Eq.~(\ref{gt_density}), since the degenerate direction $\vn$
complicates matters somewhat.

Nonetheless there are no insurmountable difficulties or dispiriting
complications here.  The main idea is that one samples a vector $\vx_\perp$
from a 3-D multivariate Gaussian in the space of vectors dual to
the dual vectors $\vw_i$ --- that is, in the 3-D vector space
of equivalence classes of vectors differing only by a multiple of $\vn$
(this is the so-called ``Quotient Space'' $V/\vn$ of the full vector space
$V$ by the subspace spanned by $\vn$). One then samples a real number
$\lambda$ from a uniform distribution in some chosen range.  The full
sample vector is $\vx=\vx_\perp+\lambda\vn+\vx_0$.

Operationally, this is straightforward.  From Eq.~(\ref{dualbasis}), it is
apparent that we can choose as representative vectors for an orthogonal
basis of the quotient space the vectors $\ve_i, i=1,2,3$.  The reduced
matrix $b^{ij}$ in Eq.~(\ref{reduced_b}) may be thought to operate on
components of vectors expressed in this basis. That is to say, we may
sample from a 3-D multivariate Gaussian with inverse covariance
components given by $b^{ij}$, ascribing the components of the sampled
vectors to the first three components of $\vx_\perp$ (whose fourth
component is zero).  One then proceeds from $\vx_\perp$ to $\vx$ as
described above.

Note that the choices $n_4=1$, $x_{0,4}=0$ imply that a vector $\vx$
sampled in this way satisfies $x_4=\lambda$.  Thus the chosen range of
$\lambda$ is also the chosen range of $x_4$.

The fact that we choose the Cholesky decomposition $L^{ij}$ to parametrize
$b^{ij}$, so that $\vb=\vL\vL^T$, is of some assistance here.  If we sample
three numbers $\vs=(s_1,s_2,s_3)$ independently from a 1-D
standard normal distribution, then the vector $(\vL^{-1})^T\vs$ is easily
seen to be sampled from a Gaussian distribution with inverse covariance
$\vb$, as required.

\subsection{Summary Of The Model}

The 4-D Gaussian Tube model is therefore characterized by 12
parameters:  6 parameters required to establish the location and
orientation of the tube through the vectors $\vn$ and $\vx_0$, and another
6 parameters required to set up the actual Gaussian distribution about that
axis, through the lower-diagonal matrix $\vL$, which is used to obtain the
quadratic form $\vB$ using Eqs. (\ref{B1}), (\ref{dualbasis}), and
(\ref{choldec}).  In a more general $N$-dimensional space of observables,
the parameter count would be $(N-1)(N+4)/2$.

Including the normalization, the expression for the model density is
\begin{eqnarray}
\rho(\vxsrc)&=&\left(\prod_{i=1}^3L^{ii}\right)\times\nonumber\\
&&
\hspace{-0.5cm}\exp\left[
-\frac{1}{2} \left(\vxsrc-\vx_0\right)^T\vB
\left(\vxsrc-\vx_0\right)
\right].\nonumber\\
\label{gt_density_norm}
\end{eqnarray}

\section{Projection\label{Sec:Proj}}

As discussed in the Introduction, we require the ability to project the
Tube onto lower-dimensional subspaces.  This may be for the sake of
visualizing the correlation in 2-D, or it may be in order to compare the
model to the data from a GRB that is not supplied with all four possible
observations.

The process of ``projecting'' a 4-D correlation down to a subspace (such as
a visualizable 2-D plane) is, in effect, marginalization over the remaining
dimensions.  This is a standard operation in Gaussian probability theory,
which will now be briefly reviewed.

Suppose, that we wish to project onto a subspace, by marginalizing the
Gaussian Tube over the complement of the subspace.  We partition all
vectors and matrices into the two subspaces:
\begin{eqnarray}
\vn=\left[
\begin{array}{c}
\vn_1\\
\vn_2
\end{array}
\right]\quad;\quad
\vx_0&=&\left[
\begin{array}{c}
\vx_{0,1}\\
\vx_{0,2}
\end{array}
\right]\quad;\quad
\vx=\left[
\begin{array}{c}
\vx_1\\
\vx_2
\end{array}
\right]\nonumber\\
\vB&=&\left[
\begin{array}{cc}
\vB_{11}&\vB_{12}\\
\vB_{12}^T&\vB_{22}
\end{array}
\right]
\end{eqnarray}
We will project onto subspace ``1'', by marginalizing over subspace ``2''.

The projected density is
\begin{eqnarray}
\eta(\vx_1)&=&\left(\prod_{i=1}^3L^{ii}\right)\times\nonumber\\
&&\hspace{-1cm}\int d\vx_2\,\exp\Bigg\{
-\frac{1}{2}\bigg[
\left(\vx_1-\vx_{0,1}\right)^T\vB_{11}\left(\vx_1-\vx_{0,1}\right)
\nonumber\\
&&
+2\left(\vx_1-\vx_{0,1}\right)^T\vB_{12}\left(\vx_2-\vx_{0,2}\right)
\nonumber\\
&&
+\left(\vx_2-\vx_{0,2}\right)^T\vB_{22}\left(\vx_2-\vx_{0,2}\right)
\bigg]\Bigg\}.
\label{proj_1}
\end{eqnarray}

The integral may be performed by completing the square.  The result is
\begin{eqnarray}
\eta(\vx_1)&=&\left(\prod_{i=1}^3L^{ii}\right)
\det|\vB_{22}|^{-1/2}\times\nonumber\\
&&\hspace{-1.cm}\exp\left[
-\frac{1}{2}\left(\vx_1-\vx_{0,1}\right)^T
\vA_{11}
\left(\vx_1-\vx_{0,1}\right)
\right],\nonumber\\
\label{proj_2}
\end{eqnarray}
where
\begin{equation}
\vA_{11}\equiv\vB_{11}-\vB_{12}\vB_{22}^{-1}\vB_{12}^T.
\label{proj_qf}
\end{equation}
The Gaussian quadratic form in the projected space is therefore $\vA_{11}$.
Note that the term $\det|\vB_{22}|^{-1/2}$ may not be dropped from the
normalization of $\eta(\vx_1)$, for the same reasons that motivate respect
for the parameter-dependence of the normalization of the full 4-D model
density $\rho(\vx)$.

It is not hard to show that $\vA_{11}\cdot\vn_1=0$, as expected.  This is
because the partitioned version of $\vB\cdot\vn=0$ is
\begin{eqnarray}
\vB_{11}\cdot\vn_1+\vB_{12}\vn_2&=&0\\
\vB_{12}^T\cdot\vn_1+\vB_{22}\vn_2&=&0,
\end{eqnarray}
so that
\begin{eqnarray}
\vA_{11}\cdot\vn_1&=&
\vB_{11}\cdot\vn_1-\vB_{12}\vB_{22}^{-1}\vB_{12}^T\cdot\vn_1\nonumber\\
&=&\vB_{11}\cdot\vn_1+\vB_{12}\vB_{22}^{-1}\vB_{22}\cdot\vn_2\nonumber\\
&=&\vB_{11}\cdot\vn_1+\vB_{12}\cdot\vn_2\nonumber\\
&=&\vB_{11}\cdot\vn_1-\vB_{11}\cdot\vn_1\nonumber\\
&=&0.
\end{eqnarray}

In summary, all there really is to know about projection is the
partitioning trick:  the projected Gaussian Tube has a direction $\vn_1$
and an offset $\vx_{0,1}$ that are merely the appropriate partitions of
their higher-dimensional counterparts, and a quadratic form given by
Eq.~(\ref{proj_qf}).

\section{Model-Data Comparison\label{Sec:ModelData}}

As was mentioned in the introduction, the comparison of model and data is
necessarily to be carried out in the observable space, and not, as is
unfortunately customary, in the source variable space.  The reason is that
this is the only sensible way to disentangle the cosmology from the data,
and permit well-defined estimation of cosmological parameters.

Suppose that the $i$th GRB (with precisely-determined redshift $z_i$)
resulted in a measurement $\vy_i$ of the event's true observables
$\vxobs$.  We will not assume that all four observables are available to
encode in $\vy_i$.  Instead, we will assume that $\vy_i$ is an
$n$-dimensional vector, with $2\leq n\leq 4$ (so, for example, if all that
is available is $\epkobs$ and $S$ then $n=2$).  We will also encode the
measurement errors of the components of $\vy_i$ as the matrix
elements of an $n\times n$ diagonal matrix $\vD_i$, defined as
\begin{equation}
[\vD_i]_{kl}=\delta_{kl}\,{\sigma_{il}}^{-2},
\end{equation}
where $\sigma_{il}$ is the measurement error on the $l$th component of
$\vy_i$.

The strategy for calculating the likelihood function of all the data is to
calculate the event likelihood $P(\vy_i|z_i,\vn,\vx_0,\vL,\mathnormal{\Omega})$. 
Since the data for different GRBs is statistically independent, the total
likelihood is the product of all the individual event likelihoods:
\begin{equation}
{\cal L}(\vn,\vx_0,\vL,\mathnormal{\Omega})=
\prod_{i=1}^N P(\vy_i|z_i,\vn,\vx_0,\vL,\mathnormal{\Omega}).
\label{total_likelihood}
\end{equation}
The problem is therefore reduced to the calculation of the event likelihood
for each GRB.

We first require the transformed model density in the full observable
space, $\xi(\vxobs)\,d^4\vxobs=\rho(\vxsrc)\,d^4\vxsrc$.  This is easily
obtained, given the redshift $z_i$, using the transformation of
Eq.~(\ref{src2obs}).  As this transformation is a pure constant offset, its
Jacobian is 1, and we have
\begin{eqnarray}
\xi(\vxobs)&=&\rho(\vxobs - \vf(z_i,\mathnormal{\Omega}))\nonumber\\
&=&\left(\prod_{i=1}^3L^{ii}\right)\,
\exp\left[
-\frac{1}{2} \Delta\vx^T\vB
\Delta\vx
\right],
\end{eqnarray}
where
\begin{equation}
\Delta\vx\equiv\vxobs-\vx_0-\vf(z_i,\mathnormal{\Omega}).
\end{equation}

If the $i$th GRB is endowed with all 4 observations, this is sufficient. 
If, on the other hand, $n<4$, we must project $\xi(\vxobs)$ down to the
appropriate space.  We adopt the partitioning ${\vxobs}^T=[\vu^T,\vv^T]$, and
project out $\vv$ to obtain $\eta(\vu)$ by the technique of
\S\ref{Sec:Proj}, obtaining
\begin{eqnarray}
\eta(\vu)&=&\left(\prod_{i=1}^3L^{ii}\right)\,
\det|\vB_{\vv\vv}|^{-1/2}
\times\nonumber\\
&&\exp\left[
-\frac{1}{2}\Delta\vx_{\vu}^T
\vA_{\vu\vu}
\Delta\vx_{\vu}
\right],
\end{eqnarray}
where
\begin{eqnarray}
\vA_{\vu\vu}&\equiv&\vB_{\vu\vu}-\vB_{\vu\vv}\vB_{\vv\vv}^{-1}\vB_{\vu\vv}^T,\\
\Delta\vx_{\vu}&\equiv&\vu-\vx_{0,\vu}-\vf_\vu(z_i,\mathnormal{\Omega}),
\end{eqnarray}
and where the meaning of the partitioned vectors and matrices should be clear
from context.

We may now convolve this distribution with the measurement error
distribution on $\vy_i$.  This is tantamount to integrating over the entire
space $\vu$ the probability that the true value of the observables should
have been $\vu$ {\it and} that the actual measured values should have been
$\vy$.  By a routine Gaussian integration, we obtain
\begin{eqnarray}
P(\vy_i|z_i,\vn,\vx_0,\vL,\mathnormal{\Omega})&=&
\int d^n\vu\,\eta(\vu)\times\nonumber\\
&&\exp\left[
-\frac{1}{2}(\vu-\vy_i)^T\vD_i(\vu-\vy_i)
\right]\nonumber\\
&=&\left(\prod_{i=1}^3L^{ii}\right)\,
\det|\vB_{\vv\vv}|^{-1/2}\nonumber\\
&&\det|\vA_{\vu\vu}+\vD_i|^{-1/2}\times\,\nonumber\\
&&\exp\left[
-\frac{1}{2}\Delta\vy_i^T
\vQ_i
\Delta\vy_i
\right],
\label{evlik}
\end{eqnarray}
where
\begin{eqnarray}
\vQ_i&\equiv&\vD_i - \vD_i\,(\vD_i+\vA_{\vu\vu})^{-1}\vD_i,\label{evlik_q}\\
\Delta\vy_i&\equiv&\vy_i-\vx_{0,\vu}-\vf_\vu(z_i,\mathnormal{\Omega})\label{evlik_y}.
\end{eqnarray}

Eqs.~(\ref{evlik}), (\ref{evlik_q}), and (\ref{evlik_y}) are  the final
result for the event likelihood, in the case where data is incomplete and
projection is necessary.  Obviously, if the full set of observations is
available, projection is not necessary, and these formulas are to be
applied by replacing $\vA_{\vu\vu}$ by $\vB$, $\det|\vB_{\vv\vv}|$ by 1,
$\vx_{0,\vu}$ by $\vx_{0}$, and $\vf_\vu(z_i,\mathnormal{\Omega})$ by $\vf(z_i,\mathnormal{\Omega})$.

As complicated as these formulas may appear, they do not represent much of
a computational challenge, as the determinants and inverses that are
required are of symmetric, positive-definite matrices with dimensionality
less than or equal to 4.  Perhaps a slightly greater computational
challenge is the code organization required to arrange for the capability
of carrying out the projection/partition of relevant matrices and vectors
along arbitrary subsets of coordinates.  This is nonetheless a manageable
programming task.

With the event likelihood in hand, we may proceed to the calculation of the
total likelihood ${\cal L}$, as given by Eq.~(\ref{total_likelihood}).

And now, we're in business.  For example, we may simultaneously optimize
${\cal L}(G,\mathnormal{\Omega})$ (where $G$ represents the Gaussian Tube parameters)
with respect to $G$ and $\mathnormal{\Omega}$, obtaining fully internally-calibrated
point estimates of both sets of parameters, and perhaps even frequentist
confidence intervals.

We may also play Bayesian games, using some choice of prior
over the parameters to trade ${\cal L}$ in for a posterior density
distribution $P(G,\mathnormal{\Omega}|O)$, where $O$ represents the observations.  We
may then marginalize  some of the parameters to produce Bayesian confidence
regions on others.  This may require a Markov Chain Monte Carlo approach,
given the large number of parameters.  Or, we may make the approximation
that marginalization is equivalent to extremization (which is true for
Gaussian distributions)\footnote{
Note, however, that the posterior
probability density over model parameters can at best be only approximately
Gaussian, despite the Gaussian nature of the GRB density model.}
, and
calculate an approximate posterior density on the cosmology parameters
$P(\mathnormal{\Omega}|O)$ by maximizing the full posterior $P(G,\mathnormal{\Omega}|O)$ with respect
to $G$ at every value of $\mathnormal{\Omega}$.

The point is, $\cal{L}$ is a genuine likelihood --- the probability of some
data given a model --- and may be pressed into service in exactly the sort
of ways that likelihoods are normally used.  The circularity concerns
that derive from the use of ``fiducial cosmologies'' to create the
source-variable ``data'' have been short-circuited by the simple expedient
of calculating the probability of the data that is directly observed.

\section{Sanity Checking\label{Sec:Sanity}}

The program of data analysis outlined so far relies on some rather abstract
and difficult-to-visualize constructions.  It is crucial that there should
be some way to visualize the relationship between the model and the data,
both to spot possible problems and to get an intuitive feeling for the
predictive content of the model.

Once a best-fit Gaussian Tube $G$ and a best-fit cosmology (or the
concordance cosmology) $\mathnormal{\Omega}$ have been fixed, this is a straightforward
thing to do.  There are six 2-D planes that may be formed from the
4 available source variables.  The best-fit Gaussian Tube model may be
projected according to the method of \S\ref{Sec:Proj} onto each of these
planes.  The projected best-fit straight line and the $1-\sigma$ confidence
region from the projected distribution may be plotted on each plane. Each
GRB endowed with observations that are representable on some of those
planes may have those observations mapped to the appropriate plane (for
example, a GRB with measured $S$, $\epkobs$, and $\dobs$ may be mapped to
the $\xiso-\xpksrc$, $\xiso-\xdsrc$, and $\xpksrc-\xdsrc$ planes).

We finally end up with a series of six plots, each one displaying the
projected model and the mapped data.  From these plots, it should be
possible to visualize directly the properties of the various projected
aspects of the correlation model, and the extent to which the best-fit
model really respects the data.

Besides this sort of visual verification of the various 2-D
correlations against the data, there is another model verification issue
that merits consideration.  The observed redshift distribution of GRBs
drops dramatically below $z\le 1$, where most SN~Ia redshifts are found, and
extends out past $z=6$. This is an opportunity, of course, since it means
that GRB-derived confidence regions in, say, the $\mathnormal{\Omega}_M-\mathnormal{\Omega}_\mathnormal{\Lambda}$
plane cut across those derived from SN~Ia \citep{ggf_2006}, furnishing
tighter constraints on those parameters.  However, the much broader range
of GRB redshifts raises a troubling question:  Even if we find a
reasonable-seeming fit of the Gaussian Tube model's projections to the
data, how do we know whether the properties of the tube should be
considered to have evolved with redshift?  That is to say, is it reasonable
to assume, as the model does, that the correlations of GRB energetics
follow the same distributions irrespective of redshift?  If so, how do we
know?  If not, how would this affect the inferred values of the
cosmological parameters $\mathnormal{\Omega}$?

This question cannot be addressed merely by inspecting the projection plots
described above, since the redshifts are all intermingled in those plots. 
Instead, it seems advisable to adopt a model-comparison strategy, wherein
the ``vanilla'' Gaussian Tube model described above is compared to more
complicated models (via a frequentist likelihood ratio, or a Bayesian test
based on posterior odds ratios) that allow the correlation parameters to
vary with redshift.  That is, we may introduce another hierarchical level
in the model by allowing some of the parameters $G$ to be some parametrized
empirical function of $z$ (a linear function is an obvious thing to try),
and calculate the amount by which the log-likelihood (say) is improved in
this model over a model in which the $G$ are the same for all redshifts.
Significant improvements would be evidence for evolution of the
distributions.  Additionally, significant shifts in the confidence regions
in the $\mathnormal{\Omega}_M-\mathnormal{\Omega}_\mathnormal{\Lambda}$ plane in the more complicated model could
be interpreted as an indication of trouble, whereas stability of those
contours as the new parameters are varied would be a reassuring sign of
robustness of the results.

Clearly this approach entails some considerable expansion of the parameter
space.  A somewhat more modest approach, similar to the calibration
approach of \citet{basilakos_etal_2008}, is to partition the GRBs into a
small number of bins, fit them separately, and determine whether the sum of
the log likelihoods is significantly better than the log-likelihood for the
full sample. Again, any significant shifts in $\mathnormal{\Omega}_M-\mathnormal{\Omega}_\mathnormal{\Lambda}$
contours, or lack of such shifts, would be telling of the robustness of the
inferences drawn from the model.

\section{Data Requirements\label{Sec:DataRequirements}}

This paper is a ``methods'' paper, and I have not yet attempted to collect
a carefully-calibrated sample of GRB data to subject to this
analysis.\footnote{
The catalog of \citet{schaefer_2007}, with its many arbitrary
manual adjustments to compensate for missing data, is probably not adequate
for this purpose.  The small statistical errors that would result from its
large size (69 events) would probably not compensate for the large
systematic errors introduced by the data aggregation procedure.}
I can therefore not suggest precise guidelines as to how many GRBs, bearing
what kind of information, may be necessary to obtain interesting
constraints on cosmological parameters using the present methodology. 
Nonetheless the question is worth addressing, so I offer a few tentative
thoughts on the matter.

The ``vanilla'' (i.e. not redshift-dependent) Gaussian Tube model presented
here has 12 free parameters, and operates on 2 to 4 observable quantities
per GRB.  In addition, a minimally interesting cosmological model offers
two additional parameters for constraining ($\mathnormal{\Omega}_M$ and
$\mathnormal{\Omega}_\mathnormal{\Lambda}$), so that a total of 14 parameters must be managed in the
fit.

Consider the Tube parameters $G$ first.  The role of these 12 parameters is
to ensure that the 6 2-D projections of the model adequately fit the
projections of available data into those planes.  The model is more compact
than a model composed of 6 2-D Tube models (which would require 18
parameters).  Therefore, a conservative estimate of the amount of data
required to constrain the full Gaussian Tube model is the amount required
to constrain the 6 independent 2-D Tube models.  In each plane, this number
would depend on the tightness of the correlation, the size of the
measurement errors, and the dynamic range of the data.  In the case of the
original Amati relation \citep{amati_etal_2002}, with 10 constrained data
points, measurement errors in the 10-30\% range, and a dynamic range in
$\eiso$ of nearly 3 orders of magnitude, the fit parameters that resulted
had a statistical error of about 10\%.  Suppose, then, for the sake of
making a conservative estimate, that we require 15 points in each plane for
adequate constraints on $G$.  The number of events required to furnish this
information is bounded below by 15 (if all events bear all information, so
that 60 numbers are used), and above by 90 (if all points on all planes are
due to different GRBs, so that 180 numbers are used).

Turning to the cosmological parameters, one may observe that initially, the
inflation of the Tube parameter count from 3 (for a single 2-D correlation
such as the Ghirlanda relation) to 12 \textit{adds} uncertainty to the contours in
the $\mathnormal{\Omega}_M$--$\mathnormal{\Omega}_\mathnormal{\Lambda}$ plane, uncertainty which must be made up
by adding data that constrains those additional Tube parameters.  If the
data is in fact constraining on those additional parameters, then one may
expect the statistical errors on cosmological parameters to shrink roughly
as $N_{pair}^{-1/2}$, where $N_{pair}$ is the number of independent pairs
of event data (i.e. the total number of points in the 6 projected planes). 
This is the point of the exercise:  by passing to the Gaussian Tube model,
one pays a price in parameter count inflation, in the expectation that one
will reap a dividend through the larger and more informative dataset that
thereby becomes accessible.

In other words, the design of this framework requires a cost/benefit
analysis.  It is not necessarily the case that the particular choice of
observables made in this work for the sake of illustration is optimal. 
Possibly a different set, or a smaller or larger set, might be preferable. 
Much depends on visual inspection of correlations.  If one or more of the
2-D projections of the data appear not to show evidence for a strong
correlation, it may be the case that the parameters controlling the
correlation in that projected plane may be adding more noise than signal,
and it might be a good idea to change observables, or to freeze the
responsible parameters at some harmless value.  On the other hand,
reasonably clear correlations of the data in all projected planes would
constitute evidence for a good choice of observables, one which is likely
to reward the analysis with statistical errors on cosmological parameters
that are smaller in consequence of more abundant data, and that shrink
more rapidly with increasing data than would errors inferred from a
lower-dimensional model.

\section{Discussion\label{Sec:Discussion}}

It is my hope that readers are persuaded that the circularity problem
of GRB standard candle enterprise was merely a diversion, an own-goal
brought about by an unfortunate choice of space for model-data comparison,
and readily corrected by making a better choice.  The various fix-ups for
the ``problem'' that have been proposed in the literature, and which were
discussed in \S\ref{Sec:Intro}, are not merely unnecessary: by taking an
excessively conservative attitude towards parameter constraints, or
actually introducing incoherent features to their statistical model, they
almost certainly do more harm than good.

The view advocated here is that the various correlations that are discussed
in the literature must necessarily be projected aspects of a
higher-dimensional ``super-correlation''.  At its root, this is really no
more than the observation that if A is correlated with B, and B is
correlated with C, then A is necessarily correlated with C, and a
correlation structure must therefore exist in the joint space of A, B, and
C.

I cannot say at this stage what the various correlations look like in all
six 2-D planes that may be constructed from the present variables. 
However, it would be difficult to understand if they weren't about as tight
as the Amati relation, unless there is something wrong with the
Ghirlanda/Firmani/Liang-Zhang/etc. correlations, which, as I explain below,
I do not believe.  Turning this around, however, there is a very
interesting possibility:  exhibiting correlations in alternative planes ---
including some built from burst durations and afterglow break times ---
strengthens the case for the reality of {\it all} these correlations, in
the sense that it is difficult to imagine a selection effect of such
perversity that it can produce both $\epksrc$--$\eiso$ and $\tsrc$--$\dsrc$
correlations (for example).

An additional remark concerning projections seems apposite.  It is possible
to search for 2-D projections that are not necessarily along the
coordinate axes, which in some sense minimize scatter in the data.  The
Ghirlanda, Liang-Zhang, and and Firmani relations are of this character. 
All that is required is to effect linear transformations of the coordinate
axes, together with the corresponding similarity transformations on all
matrices.  One could imagine searching for the linear transformation that
makes a correlation in a 2-D projection look maximally tight. 
However, it seems to me that the motivation for doing so is not as strong
in the current picture as it once was, since all the content of these
relations is already embodied in the best-fit Gaussian Tube model. 
Certainly, the construction of such a transformation would have no effect
whatever on the likelihood function computed above, or on any of the
cosmological inferences drawn therefrom.

This remark underlines the essential fact that {\it the most suitable space
for visualizing the relationship between model and data is not necessarily
the most suitable space for analyzing that relationship}.  It was the
failure to understand this truism of data analysis that gave rise to the
circularity problem in the first place.

While the reality of these correlations has been harshly questioned
\citep{nakar_piran_2005,band_preece_2005,butler_etal_2007}, in my opinion
the assuredness of these critiques is out of all proportion to the
questionable cogency of the evidence upon which they rest.  It should be
kept in mind that in order to even observe the correlations, the essential
requirements are (a) rapid, accurate astrometry (to furnish afterglow
redshifts), and (b) accurate broadband spectroscopy (to obtain the
essential spectral fit parameters).

The critiques of \citet{nakar_piran_2005} and \citet{band_preece_2005} rely
upon the fits of {\sl BATSE} spectral data reported in
\citet{band_etal_1993}, despite the fact that {\sl BATSE} had no
afterglows.  Furthermore, as attested by columns 5, 6, and 7 of Table 4 of
\citet{band_etal_1993}, many of those spectral fits were of questionable
quality.

The critique of \citet{butler_etal_2007}, relies purely on {\sl Swift}
spectral fits, but as {\sl Swift}'s bandpass is essentially 20--120~keV,
there is no possibility of securing actual spectral fit parameters.  {\sl
BATSE}-informed priors must therefore do some extremely heavy lifting. 
Nonetheless, \citet{butler_etal_2007} find an Amati Relation correlation in
{\sl Swift} data, with the correct slope, but with the wrong
normalization.  Curiously, rather than conclude that their priors might be
exerting some uncontrolled influence, they infer instead that the
inconsistency exposes the Amati relation as being due to a somewhat
vaguely-specified instrumental selection effect.

Meanwhile, every analysis based on data from instrument complements capable
of {\it both} prompt, accurate astrometry {\it and} accurate broad-band
spectroscopy, such as from {\sl BeppoSAX} \citep{amati_etal_2002} or from
{\sl HETE} \citep{sakamoto_etal_2005} has found that with the exception of
a small number of conspicuous outliers (such as the under-luminous
GRB980425), new data invariably drapes itself across the old, known
correlations.  In addition, analysis of time-resolved spectroscopy of
selected {\sl BATSE} bursts by \citet{liang_dai_wu_2004} showed that flux
and $E_{pk}$ are Amati-correlated within the time history of each GRB. 
Finally, \citet{gng_2009}, using 12 long GRBs jointly observed by {\sl
Swift} and by {\sl Fermi}/GBM (with GBM supplying the spectroscopic
coverage), not only confirm the time-resolved GRB-personalized
mini-Amati relations of long GRBs discovered by \citet{liang_dai_wu_2004},
but also show that the normalizations of those mini-relations actually
place them on the {\sl BeppoSAX/HETE} Amati relation, with the {\sl
BeppoSAX/HETE} parameters (and, of course, the time-integrated spectral
parameters of all 12 events also fall on the {\sl BeppoSAX/HETE} Amati
relation).

This debate would appear to be over: the various long GRB phenomenological
correlations, are (except for a small fraction of conspicuous outliers)
convincingly confirmed, and appear to be manifestations of ``internal''
features of GRB emission.  They must certainly be taken seriously.  Given
that {\sl Swift} and {\sl Fermi}/GBM appear capable of producing about a
dozen joint events with the required spectral data per year
\citep{gng_2009}, and given that one may expect that a sample of GRBs of
about 150 events may make an impact on Dark Energy studies comparable to
that of SN~Ia \citep{ggf_2006}, it is possible that GRBs may be put to
useful cosmological work sooner rather than later.

\section*{Acknowledgments}

I am grateful to Don Lamb for calling my attention to the circularity issue
and suggesting that it might be factitious.  Thanks are due also to
Gabriele Ghisellini and Giancarlo Ghirlanda for fruitful discussions that
assisted the formulation of the work described here.  This work is
supported in part at the University of Chicago by the US Department of
Energy under Contract B523820 to the ASC Alliances Center for Astrophysics
Nuclear Flashes, in part under NSF award AST-0909132, and in part under
NASA award NNX09AK60G.

\bibliography{grb_cosmo}

\end{document}